\documentclass[aps,prl,reprint,twocolumn,showpacs,superscriptaddress,a4paper,floatfix]{revtex4-1} 
\usepackage{hyperref}
\usepackage{graphicx}
\usepackage{dcolumn}
\usepackage{bm}
\usepackage{latexsym}
\usepackage{amsmath}
\usepackage{natbib}
\usepackage{color}

\begin{document}

\title{Estimating Excitonic Effects in the Absorption Spectra of Solids: Problems 
and Insight from a Guided Iteration Scheme}

\author{Santiago Rigamonti}\email{srigamonti@physik.hu-berlin.de}
\affiliation{Humboldt-Universit\"at zu Berlin, Institut f\"ur Physik and IRIS Adlershof, 12489 Berlin, Germany}
\affiliation{European Theoretical Spectroscopy Facility (ETSF)}

\author{Silvana Botti}
\affiliation{Institut Lumi\`ere Mati\`ere,
  UMR5306 Universit\'e Lyon 1-CNRS, Universit\'e de Lyon, F-69622
  Villeurbanne Cedex, France} 
\affiliation{Friedrich-Schiller Universit\"at Jena, Institut f\"ur 
Festk\"orpertheorie und -optik, Max-Wien-Platz 1 07743 Jena}
\affiliation{European Theoretical Spectroscopy Facility (ETSF)}

\author{Val\'erie Veniard}
\affiliation{Laboratoire des
  Solides Irradi\'es, \'Ecole Polytechnique, CNRS, CEA-DSM, F-91128
  Palaiseau, France} 
\affiliation{European Theoretical Spectroscopy Facility (ETSF)}

\author{Claudia Draxl}
\affiliation{Physics Department, Humboldt-Universit\"at zu Berlin, Germany}
\affiliation{European Theoretical Spectroscopy Facility (ETSF)}

\author{Lucia Reining}\email{lucia.reining@polytechnique.edu}
\affiliation{Laboratoire des
  Solides Irradi\'es, \'Ecole Polytechnique, CNRS, CEA-DSM, F-91128
  Palaiseau, France} 
\affiliation{European Theoretical Spectroscopy Facility (ETSF)}

\author{Francesco Sottile}
\affiliation{Laboratoire des
  Solides Irradi\'es, \'Ecole Polytechnique, CNRS, CEA-DSM, F-91128
  Palaiseau, France} 
\affiliation{European Theoretical Spectroscopy Facility (ETSF)}

\date{\today}

\begin{abstract}
A major obstacle for computing optical spectra of solids is the lack of 
reliable approximations for capturing excitonic effects within time-dependent 
density-functional theory. We show that the accurate prediction of strongly 
bound electron-hole pairs within this framework using simple 
approximations is still a challenge and that available promising results have 
to be revisited. Deriving a set of analytical formulae we analyze 
and explain the difficulties. We deduce an 
alternative approximation from an iterative 
scheme guided by previously available knowledge, significantly improving 
the description of exciton binding energies. Finally, we show how one can 
``read'' exciton binding energies from spectra determined in the random phase 
approximation, without any further calculation. 
\end{abstract}

\maketitle


The response of materials to an electromagnetic field is a key to many 
properties and applications. In the frequency range from infrared to 
ultraviolet, the optical properties determine the color of materials, their 
ability to absorb the sunlight, and much more. They lay the ground for 
non-destructive spectroscopies such as ellipsometry, that can tell us much about 
the electronic or atomic structure of materials. However, theoretical tools 
are needed that allow one to analyze, understand and predict measured results 
and desired or undesired properties. These tools should be reliable and 
versatile, but  simple enough to be applicable to systems of fundamental or 
technological interest, that are often rather complex. One of the major 
challenges is to design approximations for the  {\it ab initio} calculation of 
optical spectra of extended systems such as solids and liquids \cite{onida02}.  

The state-of-the-art approach for the {\it ab initio} calculation of optical 
spectra consists in using the Kohn-Sham (KS) electronic structure coming from 
a density functional theory (DFT) calculation as starting point for a 
quasiparticle bandstructure calculation in the $GW$ approximation, and the 
subsequent solution of the Bethe-Salpeter equation (BSE) to account for the 
electron-hole interaction~\cite{hanke1980,bse1,bse2,bse3,onida02}. The scheme is 
successful; in particular, excitonic effects are well described. However, 
calculations are computationally demanding, because of the two-particle 
(electron and hole) nature of the problem. Alternatively, time-dependent DFT 
(TDDFT)~\cite{Runge-Gross,onida02,botti-review-07} formulates the response 
in terms of variations of local potentials that are functionals of the time-dependent 
density. This reduces the size of the problem, but raises the 
question of how to find a good approximation for the time-dependent exchange-correlation 
(xc) potential $v_{\textrm{xc}}$ and its first derivative, the
xc kernel $f_{\textrm{xc}}({\bf r}, {\bf r}', t-t') = \delta v_{\textrm{xc}}
({\bf r}, t)/\delta n({\bf r}', t') $, where $n$ is the time-dependent 
electron density. Some simple but widely used approximations such as the 
adiabatic local density approximation~\cite{alda,alda2}, that are often 
successful for finite systems and for electron energy-loss spectra, yield 
disappointing results similar to the random phase approximation 
($f_{\textrm{xc}}=0$)~\cite{rpa,botti-review-07} for absorption spectra of 
solids.  

Many works, e.g.~\cite{Nazarov2009,Nazarov2011,Trevisanutto2013,pggxc,tddft-hybrids,LRC-old}, 
try to overcome this problem. A class of successful kernels has been 
derived from the 
BSE~\cite{stubner2004,vonbarth2005,bruneval2005,Sottile2003,Adragna2003,Marini2003}. 
The nanoquanta kernel~\cite{Sottile2003,Adragna2003,Marini2003,PRL-LRC} gives results 
close to BSE ones, but with a comparable computational cost, although 
suggestions for speedups have been made 
\cite{PRB-Sottile-07}. The long-range corrected (LRC) kernel~\cite{PRL-LRC,PRB-LRC} 
$f^{LRC}_{\textrm{xc}}= -\alpha/q^2$ with the correct divergence 
for small wavevectors $q$ is a simple scalar approximation of the
nanoquanta kernel. $f_{xc}^{LRC}$, with  $\alpha$ empirically determined from 
the static dielectric constant of the crystal \cite{PRB-LRC}, works well for 
continuum excitons in semiconductors \cite{lrc_app1,lrc_app2,lrc_app3}.  
However, it fails to reproduce bound 
excitons, unless $\alpha$ is set \textit{ad hoc} to a much higher value than 
in \cite{PRB-LRC}. In this case, a transition may appear within the 
quasiparticle gap \cite{PhysRevB.68.205112,turkowski2008,turkowski2009}, but
with too high oscillator strength \cite{PhysRevB.68.205112}.

Alternatively, the so-called \textit{bootstrap} (BO) kernel ~\cite{boot} also 
has the correct $1/q^2$ behavior; the prefactor is determined self-consistently, and it 
goes beyond the scalar version. Promising results have 
been published~\cite{boot,boot-eels} for continuum and bound excitons, and the 
exciton binding energies of a range of small- and large-gap semiconductors have 
been calculated ~\cite{Yang2013,Yang2012}. However, the BO expression
has not been derived, but rather justified by observations,
and the predictive power of the approach has not yet been 
demonstrated. Indeed, as we will show below, the BO does not lead to reliable 
absorption spectra, sometimes not even qualitatively. 

The aim of this work is to elucidate the origin of the BO and of its 
shortcomings, in order to go beyond. We show that a BO like expression can 
indeed be derived, but it is slightly different from the \textit{ad hoc} 
(i.e. without derivation) one of 
~\cite{boot} and it leads to improvements, in particular for exciton 
binding energies. The computational cost can be further significantly reduced 
thanks to simple analytical formulae. In particular one can ``read'' exciton 
binding energies from results obtained in the random-phase approximation 
(RPA), without any further calculation. 

Optical spectra of solids are obtained from the imaginary part of the 
macroscopic dielectric function $\epsilon_M(\omega)$, that can be calculated 
from 
\begin{equation}
 \epsilon_M(\omega) = \frac{1}{\epsilon^{-1}_{00}(\omega)} = 1-
 v_{0}\bar\chi_{00} (\omega),
 \label{eq:epsm}
\end{equation}
where $\epsilon_{{\bf G}{\bf G}'} (\omega)$ is the ${\bf q}\to 0$ limit of the 
microscopic dielectric matrix $\epsilon_{{\bf G}{\bf G}'} ({\bf q},\omega)$
in a basis of reciprocal lattice vectors. ${00}$ indicates the head 
(${\bf G}={\bf G}'=0$) element of the matrix, $v_0$ is the long range (${\bf 
G}=0, {\bf q}\to 0$) part of the Coulomb interaction, and $\bar \chi$, the 
linear density response to the total macroscopic classical potential 
\cite{onida02}, is obtained from the matrix (in ${\bf G},{\bf G}'$) Dyson 
equation 
\begin{align}
&\bar \chi (\omega)= \bar \chi^{RPA}  (\omega)+ \bar \chi^{RPA} (\omega) 
f_{xc} (\omega)\bar \chi (\omega), \label{eq:dyson}\\
&\bar \chi^{RPA} (\omega) = \chi^0 (\omega) + \chi^0  (\omega)\bar v  \bar 
\chi^{RPA} (\omega), \label{eq:RPA}
\end{align}
with $\bar v$ the Coulomb interaction without the ${\bf G}=0$ component $v_0$, 
and $\chi^0$ the independent-particle response function. The RPA solution 
$\bar \chi^{RPA} (\omega)$ includes crystal local field effects (LFE) through 
$\bar v$. Note that $\chi^0$ is in principle the Kohn-Sham independent-particle response function. 
However, here we build $\chi^0$ with quasiparticle 
energies, e.g. from a $GW$ calculation (see Supplemental Material (SM) for details \cite{sm}). 
Hence $f_{xc}$ does not have to 
simulate the gap opening with respect to the KS gap. This often adopted 
strategy for TDDFT in solids allows one to simplify the kernel significantly, 
and is used also for the BO kernel \cite{boot}. The latter is a static matrix 
(middle term below),
\begin{equation}
 f_{xc,{\bf G}{\bf G}'}^{\rm{BO}} =\frac{ \epsilon^{-1}_{{\bf G}{\bf G}'}
 (0)v_{{\bf G}'}}{1-\epsilon_{00}^{RPA}(0)} \to \frac{1}
 {\epsilon_M(0)\chi^0_{00}(0)}.
 \label{eq:fxc-boot}
\end{equation}
Often one can consider just the head element $f_{xc,00}$ without altering 
results significantly. The BO kernel is then the last term of Eq. \ref{eq:fxc-boot}, 
and Eq. \ref{eq:dyson} for $\bar \chi_{00}$ is scalar \footnote{note 
that $\bar \chi^{RPA}_{00}$ still results from a \emph{matrix} Eq. 
\ref{eq:RPA}.}. For clarity, in the following we will work with scalar 
equations unless stated. We have performed a detailed study for a family of 
matrix kernels \cite{bo-long} and found that results for the full 
matrix $f_{xc}^{\rm{BO}}$  are similar to the present scalar version. We hence 
drop the subscripts $0$ and consider the head of $\bar \chi^{RPA}$ and 
$f_{xc}$.

In  \cite{boot} the equivalent of Eqs. (\ref{eq:epsm}), (\ref{eq:dyson})  and 
(\ref{eq:fxc-boot}) were iterated numerically to self-consistency. However, 
this can easily be avoided since Eqs. \ref{eq:epsm}, \ref{eq:dyson} and 
\ref{eq:fxc-boot} combine to a quadratic equation for $\epsilon_M(0)$ with two 
solutions 
\begin{eqnarray}
 \epsilon_M (0)&=& \frac{1}{2}\left (1 + \frac{\bar \chi^{RPA}}{\chi^0} - v 
 \bar \chi^{RPA}\right ) \nonumber\\
 &\pm& \sqrt{\frac{1}{4}\left (1 + \frac{\bar \chi^{RPA}}{\chi^0}
 - v \bar \chi^{RPA}\right )^2 -\frac{\bar \chi^{RPA}}{\chi^0}},
 \label{eq:square}
\end{eqnarray}
where all quantities are static. Only the solution with the plus (+) sign is 
meaningful, since in the limit of strong screening and neglecting LFE
it leads to the RPA solution $\epsilon^{RPA}_M\to 1-v\chi^0_{00}$ as expected.
The minus sign would lead to $\epsilon_M \to 1$. Given $\chi^0 $ and $\bar 
\chi^{RPA} $, the static $\epsilon_M(\omega=0)$ from Eq. \ref{eq:square}, and 
hence $f_{xc}^{\rm{BO}}$ from Eq. \ref{eq:fxc-boot}, are numbers that can be 
determined on a pocket calculator and then used in Eq. \ref{eq:dyson} to 
correct a given RPA spectrum for excitonic effects. We have checked that the 
converged iterative results and those of Eq. \ref{eq:square} are 
indistinguishable. The next order in the strong screening expansion of Eq. \ref{eq:square} 
yields $\epsilon_M(\omega=0) = \epsilon_M^{RPA}(\omega=0) + 1$, which 
agrees with the typical magnitude of excitonic effects 
on the dielectric constant of semiconductors (see for example Table I of \cite{PRB-LRC}). 

Bound excitons occur when ${\rm Im}\,\epsilon_M(\omega_0)$, and hence ${\rm 
Im}\,\bar\chi(\omega_0)$, is non vanishing at energies $\omega_0$
within the quasiparticle gap, where  $\chi^0$ and $\bar \chi^{RPA}$ are real. 
Since the scalar $f_{xc}$ in Eq. \ref{eq:fxc-boot} is real, the imaginary part 
of Eq. \ref{eq:dyson} is \cite{PhysRevB.68.205112}
\begin{equation}
 f_{xc}  = \frac{1}{\bar\chi^{RPA}(\omega_0)}.
\end{equation}
With Eq. \ref{eq:fxc-boot}, the position $\omega_0$ of the first excitonic 
peak inside the gap is then the implicit solution of
\begin{equation}
 \bar\chi^{RPA}(\omega_0) = \epsilon_M(0) \,\chi^0(0).
 \label{eq:binding}
\end{equation}
By plotting ${\rm Re}\,\epsilon_M^{RPA}(\omega)$ and comparing to the 
static $1- v\epsilon_M\chi^0$ with $\epsilon_M$ from Eq. \ref{eq:square}, 
one can hence ``read'' exciton binding energies from an RPA spectrum.
For illustration, we show bulk silicon, LiF and solid argon. 
\begin{figure}[!htbp]
\centering
   \includegraphics[width=1.0\columnwidth]{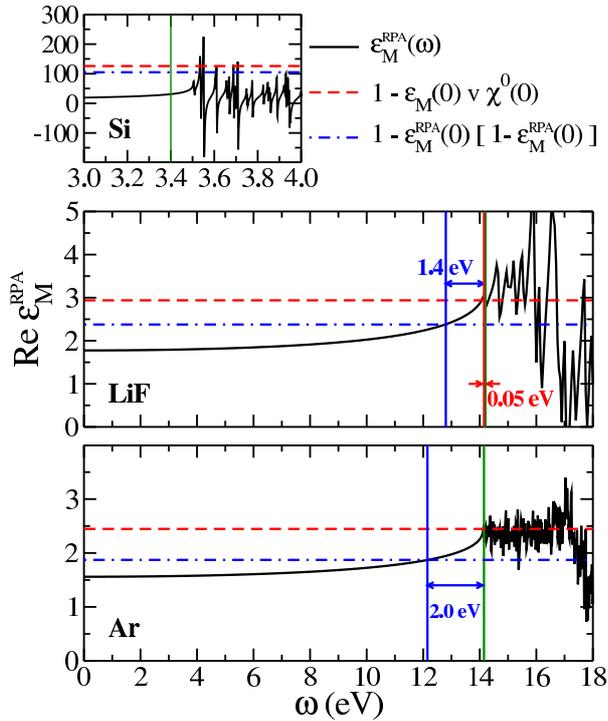}
\caption{
(color online) Real part of $\epsilon_M^{RPA}(\omega)$ for Si, LiF, and Ar
(black solid line). Its crossing with the red dashed (blue dot-dashed) 
horizontal lines gives the exciton binding energy $\omega_0$ from Eq. 
\ref{eq:binding} (\ref{eq:bindingRBO}). The green vertical line indicates the
quasiparticle gap.
}
\label{fig:binding-energies}
\end{figure}
The black solid lines in Fig. \ref{fig:binding-energies} show the real part
of $\epsilon_M^{RPA}(\omega)$ for the three materials (for computational
details, see the SM \cite{sm}); it is monotoneously increasing
within the quasiparticle gap. The value $[1-\epsilon_M(0)v\chi^0(0)]$ with 
$\epsilon_M$ calculated with Eq. \ref{eq:square} is given by the horizontal
red dashed lines, and the red vertical lines indicate intersections, hence 
bound excitons.
 
In silicon no bound exciton is found because $\epsilon_M(0)$ is large. 
LiF and argon have a low dielectric constant and therefore exhibit a crossing
below the gap. However, the exciton binding energies, given by the difference
between the energy of the fundamental quasiparticle gap and the exciton peak, 
are only 0.05 eV in LiF and 0.0 eV in Ar, much smaller than the experimental 
results (about 1.4 eV and 2.0 eV, respectively \citep{roessler,saile2}), and
in apparent contrast to \cite{boot}. The latter discrepancy  cannot be 
explained with the use of Eq. \ref{eq:binding}, which is exact when the BO 
kernel is used. Let us therefore look at the spectra.
\begin{figure}[!htbp]
\centering
   \includegraphics[width=1.0\columnwidth]{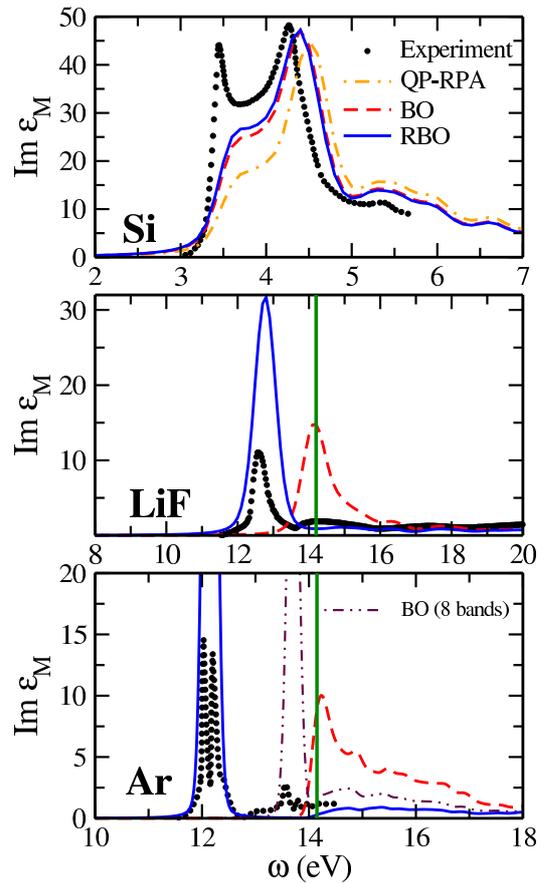}
\caption{
(color online) Imaginary part of $\epsilon_M^{RPA}(\omega)$ for Si, LiF, and 
Ar computed in various approximations. Experimental spectra are taken from \cite{cardona} for Si, 
\cite{roessler} for LiF and \cite{saile2} for Ar. The green vertical lines 
indicate the quasiparticle gap.}
\label{fig:si_lif_ar}
\end{figure}
Fig. \ref{fig:si_lif_ar} shows our results of BO calculations for the 
imaginary parts of the macroscopic dielectric function for Si, LiF and Ar.
Red dashed curves stem from our TDDFT calculations with the BO kernel (Eq. 
\ref{eq:fxc-boot}). 
In silicon, like in \cite{boot}, the kernel improves the spectrum with respect 
to QP-RPA \footnote{QP-RPA stands for an RPA calculation with $\chi_0$ built with quasiparticle energies} by enhancing the first peak and inducing an overall transfer of 
oscillator strength to lower energies. However, the effect appears 
underestimated when compared to experiment \cite{cardona} and to BSE \cite{PRB-LRC}. 
For silicon, the simple long-range $f_{xc}^{LRC}=\alpha/q^2$ is sufficient 
\cite{PRL-LRC}. However the prefactor $\alpha_{\rm{BO}}$ from the BO kernel 
defined in Eq. \ref{eq:fxc-boot}, is only $\alpha = -0.1$, too weak compared 
to the optimal value $\alpha = -0.2$ \cite{PRB-LRC}, which explains why the 
effect is not strong enough.

In LiF and Ar, our BO spectra confirm the weak exciton binding energies 
obtained from Eq. \ref{eq:binding}. The  spectral shapes look similar to the 
ones of \cite{boot}; however, the positions of the exciton peak differ and, for 
argon, the peak height from our BO is about half of that in \cite{boot}. As 
regards the peak position, we are not in contradiction with \cite{boot} since 
our quasiparticle gaps are close to experimental 
photoemission gaps. Instead, the quasiparticle gaps used in \cite{boot}
are much smaller. This compensates the too small exciton binding energy and 
leads to seemingly good agreement with experimental optical spectra. 
We elaborate on this point in the SM \cite{sm}. Additionally, it is important to note
that the exciton binding energy is very sensitive to details, especially for strongly bound
excitons. The reason is that the latter lie in a region where the real part of
$\epsilon^{RPA}(\omega)$ is very flat (see Fig. \ref{fig:binding-energies}).
A small change in $\chi^0(0)$ leads then to a large shift in the crossing
point, and hence in the exciton binding energy. Such a small change in $\chi^0(0)$
can be due to a small change of the structure, or of computational ingredients like a
pseudopotential or convergence parameters, and it can be amplified
since the static dielectric constant enters the BO self-consistently.
Indeed, the calculations for the BO kernel show a notable slow convergence
with respect to both the LFE (i.e., number $N_G$ of ${\bf G}$ vectors) and
the number of empty bands. The second issue is exemplified in Fig.
\ref{fig:si_lif_ar} for the case of Ar: the brown double-dot-dashed curve has
been obtained with only 8 bands, versus 20 in the converged calculation 
(red dashed curve). The unconverged calculation exhibits a bound exciton with a binding
energy of more than half an eV. Similarly, poorly converged calculations with respect to
$N_G$ give also, for the case of argon, a BO spectrum with a slightly higher
binding energy and a higher
peak height than the converged result \cite{bo-long}, much more similar to
\cite{boot}. More generally, this explains why for argon or LiF one
can easily obtain results that differ by an eV or more from others in the
literature \cite{Yang2013}.

Once the calculations are settled, the results of the BO are hence 
disappointing. Let us therefore finally elucidate the origin of this kernel 
and indicate a possible improvement. We start from three assumptions:
\begin{itemize}
 \item [A] We can take a static $f_{xc}(\omega=0)$ in the optical range. 
 \item [B] The static dielectric constant is larger than one.
 \item [C] The static dielectric constant is not too different from
   the RPA one.
\end{itemize}
These assumptions are based on previous knowledge from theory and experiment 
(B), numerical results e.g. of Bethe-Salpeter calculations (C), and, most 
importantly (A), insight from previous studies of long-range corrected 
kernels, e.g. \cite{PRL-LRC,PRB-Sottile-07,delsole03}. The fact that $f_{xc}$ 
should be proportional to the inverse dielectric constant \cite{PRL-LRC,PRB-LRC,scr_exch1,scr_exch2} 
has also been useful to guide the derivation, which we start by combining 
Eqs. \ref{eq:epsm} and \ref{eq:dyson}:
\begin{equation}
 f_{xc} = \frac{1}{\bar \chi^{RPA}} - \frac{1}{\bar \chi} = \frac{1}{\bar 
 \chi^{RPA}} - \frac{v}{1-\epsilon_M}.
\label{eq:fxc-chi}
\end{equation}
If one had to make a guess for $f_{xc}$ and iterate Eqs.\ref{eq:epsm}, 
\ref{eq:dyson}, and \ref{eq:fxc-chi}, one would of course get the same $f_{xc}$ 
back, however absurd it might be. The trick of a BO-like approach is to make 
an approximation in one of the equations, such that they are no longer 
equivalent. At first sight this should not lead to any advantage: how could an 
approximation be better than the exact formula? However, by choosing the 
approximation carefully one can feed information. In that case, iteration of 
the (now no longer equivalent) equations may indeed define the three unknowns 
$f_{xc}$, $\bar \chi$ and $\epsilon_M$. We will call this procedure ``guided 
iteration''.  We will first use condition (A) to this aim: a static kernel can 
be determined from the equations at $\omega=0$ alone. In that limit, 
hypothesis (B) is generally valid, and we can use it to expand the $1/\bar 
\chi$ term in Eq. \ref{eq:fxc-chi} to leading order in $1/\epsilon_M$,
\begin{equation}
 \frac{1}{\bar \chi} = \frac{v}{1-\epsilon_M} \approx - \frac{v}{\epsilon_M} 
 \approx - \frac{v}{\epsilon^{RPA}_M}
\label{eq:approx}
\end{equation}
where we have used  hypothesis (C) in order to obtain the last expression.
This finally leads to
\begin{equation}
  f_{xc}^{RBO} \approx \frac{1}{\epsilon_M^{RPA}\bar \chi^{RPA}},
\label{eq:bs1}
\end{equation}
which we call RBO (RPA bootstrap). The RBO is close to the BO which appears in Eq. \ref{eq:fxc-boot}, but there is no self-
consistency condition. The blue curves in Fig. \ref{fig:si_lif_ar} are 
obtained using Eq. \ref{eq:bs1}.  In silicon, the 
improvement with respect to the RPA result is close to that of the original BO 
(red dashed curve). Changes are noticeable in LiF and argon, where now the 
peak position is close to the experimental one \cite{so_ar}. Compared to 
experiment there is still too much spectral weight on these peaks. This is to be 
expected, because the two 
kernels behave like the LRC. It is indeed known \cite{PhysRevB.68.205112} that 
one can tune $\alpha$ to reproduce the exciton binding energy, but at the 
price of too much oscillator strength. To cure this problem, one may have to 
introduce a frequency dependence that is able to distribute spectral weight 
over the whole Rydberg series; this is however beyond the scope of the present 
work. Here we focus on the exciton binding energy, that can now again be 
obtained from $\epsilon_M^{RPA}(\omega)$ alone, using the modified 
prescription
\begin{equation}
  \bar\chi^{RPA}(\omega_0) = \epsilon_M^{RPA}(\omega = 0) \bar \chi^{RPA}
  (\omega = 0).
  \label{eq:bindingRBO}
\end{equation}
This corresponds to the use of the blue horizontal dot-dashed line in Fig. 
\ref{fig:binding-energies}. The exciton binding energies that we can read in 
this way, and that correspond of course to the peak positions given by the 
RBO in Fig. \ref{fig:si_lif_ar}, are 2.0 eV for argon and 1.4 
eV for LiF, in excellent agreement with the experimental values of 2.0 eV and 
1.43 eV, respectively.

In conclusion, starting from the so-called bootstrap kernel of TDDFT 
\cite{boot}, we have derived very simple approaches to determine absorption 
spectra and to estimate exciton binding energies from RPA calculations alone. 
We have however shown that the boostrap kernel is not reliable for the  
determination of exciton binding energies, and that promising results in the 
literature are partially misleading. We have therefore derived a related 
kernel starting from a few physically meaningful assumptions. Numerical 
results confirm that the new kernel is more reliable. One may expect that this 
first derivation of a bootstrap-like kernel could trigger new developments, 
but caution is called for: our ``guided iteration'' is not a systematic 
expansion that one might continue to obtain better and better results, since 
it intrinsically relies on the fact that an approximation is made by 
feeding 
knowledge. We stress again 
the importance of this approximation: 
Eqs. (\ref{eq:epsm})-(\ref{eq:dyson}) and Eq. (\ref{eq:fxc-chi}) are equivalent, though 
written in a 
different way; however making an approximation on the second term of Eq. 
(\ref{eq:fxc-chi}) [as in Eq. (\ref{eq:approx})] leads to a new formula, so breaking
the otherwise tautological sequence 
Eqs. (\ref{eq:epsm})-(\ref{eq:dyson}) and Eq.(\ref{eq:fxc-chi}). The choice of a 
reasonable approximation (in this case the RBO) makes the method 
very effective in the description of the spectrum and, above all, 
for estimates of exciton binding energies. As we have 
shown and explained, these estimates 
are very sensitive, and a numerically precise 
agreement should not be overemphasized. Most importantly, we have shown 
that exciton binding energies can be obtained at literally zero cost, 
since we have introduced a way to read binding energies from RPA dielectric 
functions alone. This may be interesting especially for scientists outside the 
community of {\it ab initio} calculations, including experimentalists, since 
it allows one to use the numerous already published RPA results, without the 
need of new calculations. 

\begin{acknowledgments} 
We are grateful for helpful discussions with S. Sharma and C. Ullrich.
We acknowledge support by the French ANR (projects NT09-610745 and ANR-12-BS04-0001-02),
the Einstein Foundation Berlin, and the DFG (Deutsche Forschungsgemeinschaft). 
Computer time was granted by GENCI (544). 
\end{acknowledgments}

\bibliographystyle{apsrev4-1}

\begin{thebibliography}{62}
\bibitem [{so_()}]{so_ar}%
  \BibitemOpen
  \href@noop {} {}\bibinfo {note} {Note that our calculations do not include
  the spin-orbit coupling, which is experimentally observed in Ar.}\BibitemShut
  {Stop}%
\end{thebibliography}

\end{document}